\documentclass[prd,aps,12pt]{revtex4}
\usepackage{amssymb,amsmath,amsthm}
\usepackage{mathrsfs}
\usepackage[applemac]{inputenc}
\usepackage{graphicx}

\begin{document}

\title{Accelerating universe as a result of an adjustment mechanism\footnote{
Essay awarded honorable mention in the Gravity Research Foundation essay competition 2015.}}

\author{Prado Mart\'in-Moruno\footnote{Present address: Departamento de F\'isica Te\'orica I, Ciudad Universitaria,
Universidad Complutense de Madrid, E-28040 Madrid, Spain.}}
\email{pmmoruno@fc.ul.pt}
\author{Nelson J.~Nunes}
\email{njnunes@fc.ul.pt}
\affiliation{Instituto de Astrof\'isica e Ci\^encias do Espa\c{c}o, Universidade
 de Lisboa, Faculdade de Ci\^encias, Campo Grande, PT1749-016 Lisboa, Portugal}


\begin{abstract}
In this essay we propose that the theory of gravity's vacuum is described by a de Sitter geometry.
Under this assumption we consider an adjustment mechanism able to screen any value of the vacuum energy of the matter fields.
We discuss the most general scalar-tensor cosmological models with second order equations of motion that have a fixed de Sitter critical point for any kind of material content. 
These models give rise to interesting cosmological evolutions that we shall discuss.
\end{abstract}

\pacs{04.50.Kd, 95.36.+x, 98.80.Jk}
\keywords{dark energy, alternative theories of gravity, cosmology}

\maketitle

\section{Introduction}

As is well known, vacuum energy gravitates in general relativity.
Assuming that quantum field theory is valid up to the Planck scale, one can estimate that this energy is of order $M_{\rm Pl}^4\sim10^{72} {\rm GeV}^4$.
If the effect of such a huge vacuum energy is not cancelled by the bare cosmological constant and other vacuum contributions,
it will dominate the late time cosmological dynamics leading to a universe in which galaxies would have never been formed. 
The fine-tuning needed to produce such a  cancellation is the first part of the cosmological constant problem \cite{Weinberg:1988cp,Carroll:2000fy}.
The second part is that this fine-tuning has to be carried out successively every time a cosmological phase transition changes the value of the cosmological constant or due to radiative corrections 
 \cite{Kaloper:2014dqa}. 

The evidence for the current accelerated expansion of the Universe has revived this problem. In the framework of general relativity and assuming the validity of the energy conditions \cite{H&E}, this
accelerated expansion implies the need for an observable cosmological constant $120$ orders of magnitude smaller than $M_{\rm Pl}^4$.
An alternative approach consists of considering modifications to the theory of gravity to describe this late time acceleration. 
Nevertheless, if vacuum energy gravitates also in the new theory and if we are still unable to construct an adjustment mechanism capable of screening different values of the vacuum energy, 
the cosmological constant problem persists \cite{Weinberg:1988cp}.
Thus, one usually assumes that an unspecified symmetry or mechanism evades this problem.

The simplest extension of general relativity consists of a four-dimensional metric theory of gravity with an extra degree of freedom described by a scalar field.
The Lagrangian of this new theory is constrained by the requirement that the Hamiltonian is bounded from below in order to avoid the Ostrogradski instability \cite{Woodard:2006nt}. 
For theories with up to second-order field equations, this condition is indeed satisfied.
Nonetheless, Nicolis {\it et al.}~\cite{Nicolis:2008in} have shown that consistent theories can be formulated in flat space even for a Lagrangian containing second order field derivatives. 
The field, however, must admit a  galileon symmetry that reduces to a shift-symmetry in the covariantized Galileons \cite{Deffayet:2009wt}.
More recently, Deffayet {\it et al.}~\cite{Deffayet:2011gz} generalised the covariantized Galileons formulating a theory which turns out to be equivalent to the Horndeski theory. 
This is the most general scalar tensor theory of gravity with second order field equations, found by Horndeski in 1973 \cite{Horndeski:1974wa}.

The Horndeski Lagrangian is greatly simplified in cosmological scenarios, where the metric takes the FLRW form with scale factor $a(t)$, spatial curvature $k$, and homogeneous scalar field, $\phi(t)$.
The minisuperspace Lagrangian can be expressed, up to total derivatives, in terms of $\phi$ and $\dot\phi$ only \cite{Charmousis:2011bf,Charmousis:2011ea},
\begin{equation}\label{Lsimple}
 L_ {\rm H}\left(\phi,\,\dot\phi,\,a,\,\dot a\right)=a^3\sum_{i=0}^3 \left[X_i\left(\phi,\,\dot\phi\right)-\frac{k}{a^2} Y_i\left(\phi,\,\dot\phi\right)\right] \,H^i,
\end{equation}
with $ L_ {\rm H}=V^{-1}\int {\rm d}^3 x\,\mathcal{L}_H$ 
where $V$ is the spatial integral of the volume element, $H=\dot a/a$ is the Hubble parameter, a dot represents a derivative with respect the cosmic time $t$  
and the functions $X_i$ and $Y_i$ are given in terms of the Horndeski functions \cite{Charmousis:2011ea}.

In this general framework, Charmousis {\it et al.}~\cite{Charmousis:2011bf,Charmousis:2011ea} avoided Weinberg's no-go theorem \cite{Weinberg:1988cp} regarding the absence of an adjustment mechanism 
able to screen any value of the vacuum energy, by relaxing one of the assumptions. More specifically, 
they allowed the field to acquire a non-trivial time dependence once the screening takes place.
This particular adjustment mechanism is very robust and also screens the effect of any kind of matter content.  
The crucial point is that they required the adjustment to be dynamical to allow for a non-trivial cosmological evolution; thus, Minkowski is a critical point of the dynamics.
Following this reasoning, Charmousis {\it et al.}~obtained the most general scalar field cosmological models able to self-adjust and called them the {\it fab four}, since the models are described by four independent functions of the field.
The effect of each of the four functions away from the critical point is equivalent to the effect produced by a stiff fluid, radiation or curvature \cite{Copeland:2012qf}.
When the critical point is an attractor, the late time screening may alleviate the cosmological constant problem cancelling the effect of the vacuum energy when it is about to affect the dynamics. 

This adjustment mechanism, however, forces the universe to decelerate its expansion while approaching the Minkowski final state. 
Thus, a late time accelerating cosmology does not naturally arise suggesting that self-adjustment is incompatible with the late time acceleration. 
Nevertheless, this is not necessarily the case when the self-adjustment concept is extended to non-Minkowskian final states. 
This possibility was first explored in reference \cite{Martin-Moruno:2015bda} showing that the current accelerated expansion could be a by-product of this adjustment mechanism.

\section{Horndeski fields self-adjustment to de Sitter}

It is commonly assumed that a theory of gravity's vacuum (in the absence of particle physics) must be Minkowski space. This prejudice stems from the need to avoid potential issues with field theory and to impose unitarity of the $S$ matrix.
Nonetheless, it is fair to wonder whether this  requirement is unavoidable. 
In this context, we consider an adjustment mechanism screening the vacuum energy of matter fields leading to a spatially flat de Sitter space. 
Mart\'{\i}n-Moruno {\it et al.}~\cite{Martin-Moruno:2015bda,Martin-Moruno:2015lha} have shown that if the field self-adjusts to a de Sitter vacuum, 
the late time accelerated expansion of the universe can be naturally described as the approach to a stable critical point of the dynamics. 
The value of $\Lambda$ characterizing the attractor is, therefore, unrelated to  the vacuum energy of the matter fields.

To explore this possibility we assume a spatially flat FLRW, and a fluid minimally coupled to the metric and non-interacting with the field.
We make use of the three self-adjustment conditions described in Charmousis {\it et al.}~\cite{Charmousis:2011bf,Charmousis:2011ea}, 
which were established for a screening leading to a late-time Minkowski space, and we apply them here to a late-time de Sitter space \cite{Martin-Moruno:2015bda}. 
Firstly, the field equation has to be trivially satisfied at the critical point, leading the field free to self-adjust. 
Secondly, the Hamiltonian at the critical point must depend on $\dot\phi$, so that the continuous field can absorb discontinuities on the vacuum energy. 
Thirdly, the full scalar field equation of motion must  depend on $\ddot a$, allowing a non-trivial cosmology to occur before the screening takes place.

It can be seen that the Lagrangian density evaluated at the spatially flat de Sitter critical point has the simple form \cite{Martin-Moruno:2015bda}
\begin{equation}\label{Lon}
 \mathcal{L}_{\rm H}^{\rm c}=3\sqrt{\Lambda}\,h(\phi)+\dot\phi\, h_{,\phi}(\phi),
\end{equation}
where $h(\phi)$ is an arbitrary function and {\it c} means evaluated at the critical point. 
The cosmological models capable to self-adjust to de Sitter can be classified in two different families, that we denote as the {\it Magnificent seven} and
{\it their nonlinear friends}.

\subsection{The Magnificent seven}

Let us consider the subset of models where  the $X_i$ functions in Lagrangian (\ref{Lsimple}) with $k =0$ are given by
\begin{equation}\label{ms}
 X_i^{\rm m7}=3\sqrt{\Lambda} \,U_i(\phi)+\dot\phi\,W_i(\phi),
\end{equation}
where the potentials $W_i(\phi)$ and $U_i(\phi)$ must satisfy the constraint 
\begin{equation}
\label{conditionl}
\sum_{i=0}^3W_i(\phi)\Lambda^{i/2}=\sum_{j=0}^3U_{j,\phi}(\phi)\Lambda^{j/2},
\end{equation}
such that (\ref{Lon}) in valid.
As there are eight of these functions and only one constraint, there are in fact only seven independent functions. These are the {\it Magnificent seven}.

\begin{figure}[t]
\centering
\includegraphics[width=0.49\textwidth]{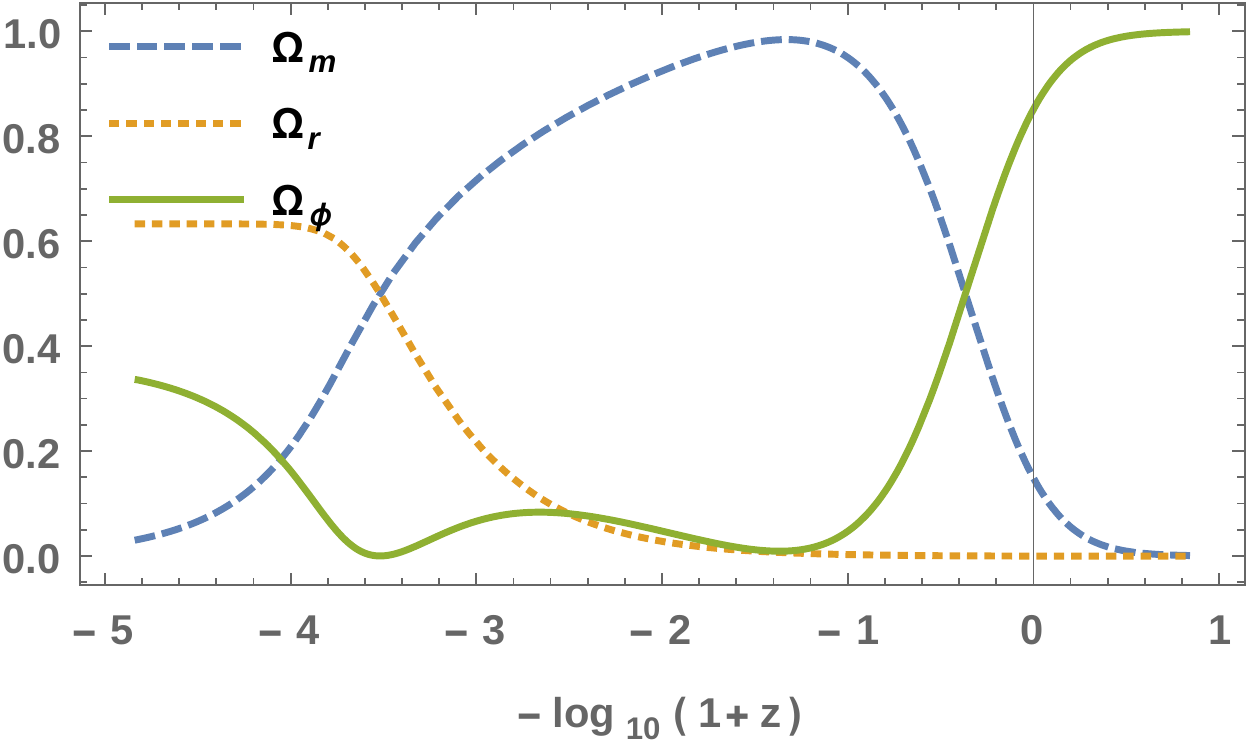}
\includegraphics[width=0.49\textwidth]{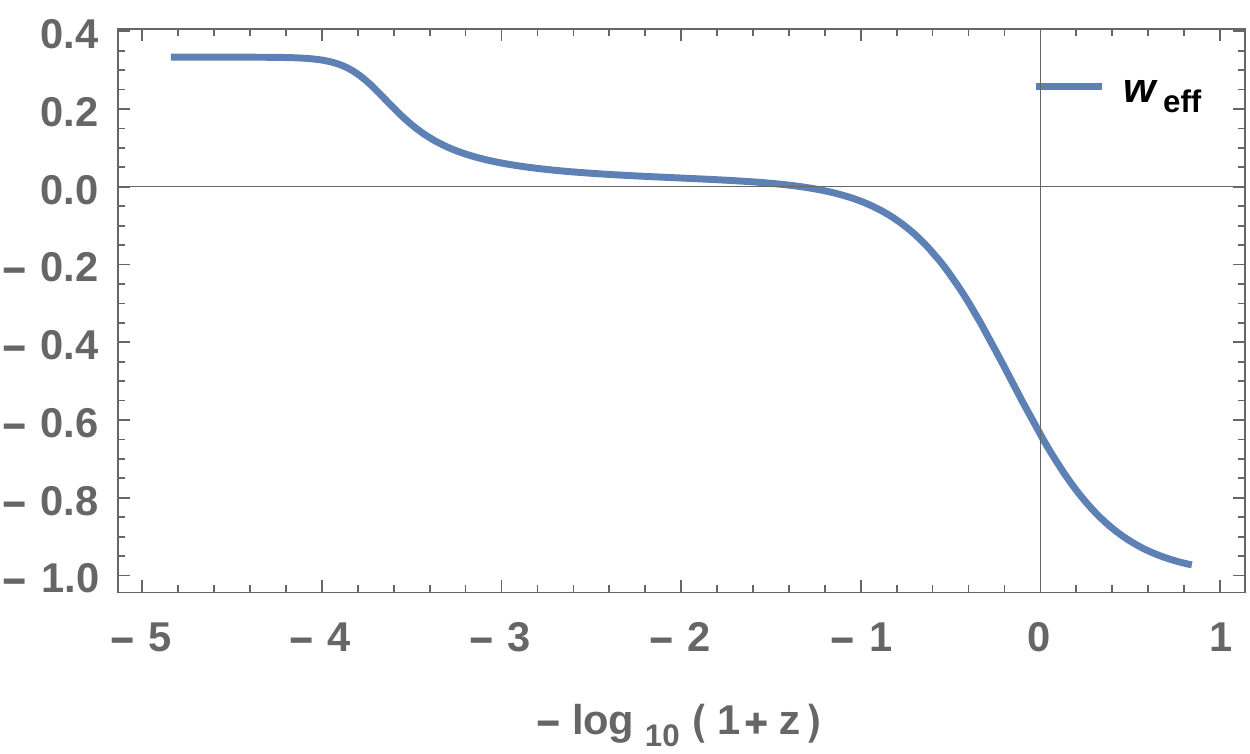}
\caption{Evolution of the abundances and of the effective equation of state parameter for the tripod model discussed in the text with $\lambda=10$, $\beta=2$ and $\gamma=-0.01$. }
\label{Fig1}
\end{figure}

The simplest and phenomenologically interesting models consist of three non-vanishing potentials  related by the condition above. 
A particularly promising example was given in reference \cite{Martin-Moruno:2015lha} by
\begin{eqnarray}
\label{trippot}
 U_2 &=&-\frac{1}{8\pi G\sqrt{\Lambda}}+\frac{3}{8\pi G \sqrt{\Lambda}}\left(e^{\lambda\phi}+\frac{4}{3}e^{\beta\phi}\right),\\
 W_2&=&\frac{3}{8\pi G \sqrt{\Lambda}}\left(\lambda e^{\lambda\phi}+\beta e^{\beta\phi}\right),
\end{eqnarray}
and with $U_3$ determined by equation (\ref{conditionl}) up to a constant $\gamma$, 
for which the evolution of the relevant cosmological parameters is illustrated  in figure \ref{Fig1}. 
Even though the qualitative good features of a viable model are present, it is clear that the field contribution is too large at present to be compatible with current observational bounds. 
In order to avoid this problem we need some help from their {\it nonlinear friends}.

\subsection{The nonlinear friends}
The $X_i$-terms of Lagrangian (\ref{Lsimple}) can have a nonlinear dependence on $\dot\phi$ provided
their contribution vanishes at the critical point and (\ref{Lon}) is recovered. Thus, the nonlinear models must satisfy
\begin{equation}\label{nl}
 \sum_{i=0}^{3} X^{\rm nl}_i\left(\phi,\,\dot\phi\right)\Lambda^{i/2}=0.
\end{equation}
There are only three independent terms but they are functions of both $\phi$ and $\dot\phi$, thus we have much more freedom than with the Magnificent seven.

For the case of a shift-symmetric field, i.e., $\phi$ independent models, and considering an Einstein Hilbert term (which does not spoil screening),  
the field equations can be combined with the conservation of the matter content to yield a closed autonomous system \cite{progress}. 
One can then verify that the critical point is indeed an attractor if the matter content satisfies $w>-1$.
This fact together with the non-renormalizability of the shift-symmetric terms augurs a possible resolution of the cosmological constant problem leading to a consistent cosmological history \cite{progress}.

\section{Conclusion}
The general and robust adjustment mechanism here described opens up the possibility to construct theories of gravity for which the vacuum energy does not gravitate when it becomes important in the late cosmological evolution. 
We have  given explicit examples of promising models and have set the framework to consider non-linear models under a shift-symmetry of the field. The investigation of the cosmology of these  models is now imperative given the pressure to 
understand theoretical models concerning dark energy  in time for them to be confronted by current and forthcoming data.


\begin{acknowledgments}
The authors are grateful to Francisco S.~N.~Lobo for collaboration in previous works as well as for encouragement. 
The authors acknowledge financial support from  Funda\c{c}\~{a}o para a Ci\^{e}ncia e Tecnologia under the 
grants EXPL/FIS-AST/1608/2013 and OE/FIS/UI2751/2014.
\end{acknowledgments}


\end{document}